\documentclass[10pt,letterpaper]{article}
 \pdfoutput=1
\usepackage[top=0.85in,left=1in,footskip=0.75in]{geometry}
\usepackage{amsmath,amssymb}
\usepackage[utf8x]{inputenc}
\usepackage{cite}
\usepackage[table]{xcolor}
\usepackage{array}
\usepackage{graphicx}
\usepackage{url}
\usepackage{authblk}

\newcolumntype{+}{!{\vrule width 2pt}}

\DeclareMathOperator{\DegC}{^\circ \mathrm{C}}

\begin{document}

\title{Long-term temporal evolution of extreme temperature in a warming Earth}

\author[1,2]{Justus Contzen}
\author[3]{Thorsten Dickhaus}
\author[1,2]{Gerrit Lohmann}
\affil[1]{Section Paleoclimate Dynamics, Alfred Wegener Institute Helmholtz Centre for Polar and Marine Research, Bremerhaven, Germany}
\affil[2]{Department of Environmental Physics, University of Bremen, Bremen, Germany}
\affil[3]{ Institute for Statistics, University of Bremen, Bremen, Germany}

\maketitle

\section{Abstract}
We present a new approach to modeling the future development of extreme temperatures globally and on a long time-scale by using non-stationary generalized extreme value distributions in combination with logistic functions. This approach is applied to data from the fully coupled climate model AWI-ESM. It enables us to investigate how extremes will change depending on the geographic location not only in terms of the magnitude, but also in terms of the timing of the changes. We observe that in general, changes in extremes are stronger and more rapid over land masses than over oceans. In addition, our models differentiate between changes in mean, in variability and in distributional shape, allowing for developments in these statistics to take place independently and at different times. Different models are presented and the Bayesian Information Criterion is used for model selection. It turns out that in most regions, changes in mean and variance take place simultaneously while the shape parameter of the distribution is predicted to stay constant. In the Arctic region, however, a different picture emerges: There, climate variability drastically and abruptly increases around $2050$ due to the melting of ice, whereas changes in the mean values take longer and come into effect later.

\section{Introduction}
\label{sec:introduction}

In many regions of the world, a rising trend in frequency and magnitude of temperature extremes is currently observed (\cite{Rahmstorf, lorenz, ipcc_ar6}). Heatwaves and extreme temperatures can have devastating effects on human societies and ecosystems (\cite{basaganha,Stillman,hatfield}) as well as on economies and agriculture (\cite{miller, Garcia-Leon}). The consequences of an increase in extreme events can also be considerably more severe than those of changes in mean temperature alone (\cite{germain}), explaining why the investigation of future changes in climate extremes is an increasingly active research topic (\cite{morak, oneill}). However, the long-term development of temperature extremes on a global level is still less well understood, and the focus of most studies is on regional investigations or on the near future (\cite{rummukainen}). It has been observed that changes in extreme temperature are not taking place uniformly around the globe, but that they are instead showing a strong dependency on the geographic location and its climatic conditions. This is visible already on a regional level (\cite{charlotte, twardosz}) and even more globally (\cite{trenberth_jones}). Model simulations predict this variability also to be present in the future development of Earth's climate (\cite{li, darand}). Changes in extremes events can be caused by changes in various statistical parameters, like the mean and the variance (\cite{parey}). In addition to that, developments can also vary temporally in different regions. In our work, we create mathematical models to investigate changes in temperature extremes in a warming climate on a global scale and for a long period of investigation.  In order to get insights into the questions outlined above, the models will be applied to temperature data from the global fully-coupled climate model Alfred Wegener Institute Earth System Model (AWI-ESM) consisting of a historical run and a future simulation and ranging from $1850$ to $2300$. \\

It is expected that the rate of change of extremes will increase in the near future (\cite{smith}). Under the premise that mankind will be able to slow and ultimately end the increase of atmospheric CO\textsubscript{2} emissions someday, it can be expected that in consequence, changes in extreme temperatures will gradually slow down as the climate system will be tending toward a new equilibrium state (\cite{king}), although it may still take centuries for a new stationary state to be completely reached due to slow-changing components of the climate system (\cite{hansen}). Taking these considerations together, we can expect changes in extreme temperature to follow in general a slow \textendash{} fast \textendash{} slow pattern over time. To describe a transition from an initial value to a final one that starts slowly, then speeds up and finally decelerates again when approaching the new value, it is common practice to use a logistic function, which exhibits a characteristic S-shaped form. The first application of logistic functions in modeling is due to Verhulst, who designed a logistic growth model to describe the development of biological populations in 1845 (\cite{Verhulst1845}). The motivation in the ecological context is that the population growth is slow at the beginning (limited by the small population size) as well as at the end (limited by the lack of natural resources). The logistic growth model has been successfully applied in biology and epidemiology  \textemdash{} a recent example being its application to the COVID-19 disease (\cite{shen})  \textemdash{} and this has motivated its use as a general model to describe changes from one state to another in fields as varied as linguistics (\cite{altmann}), medicine (\cite{yano}) or economics (\cite{kwasnicki}). \\

The analysis of extremes is complicated by the fact that extreme events are often rare, and that it is therefore difficult to build informative statistics based solely on the extreme events themselves. One common approach to overcome this issue is based on block-wise maxima: The data are split up into different (time) blocks of a sufficiently large size and then the maxima of each block are investigated. Under suitable conditions, the block-wise maxima can be approximated with a generalized extreme value (GEV) distribution (\cite{fisher_tippett, gnedenko}). GEV distributions have found numerous applications in climatology and hydrology, examples include \cite{sarr}, \cite{najafi} and \cite{chu}. A GEV distribution is determined by three parameters, called "location", "scale" and "shape", with the latter one describing the heavy-tailedness of the distribution. To model extremes in a changing climate, we will use non-stationary GEV distributions with time-dependent distribution parameters. The changes in the distribution parameters will be described using logistic functions. After fitting the models to the data, we will analyze the estimated distribution parameters in detail, and we will use the estimates also to investigate futue changes in the distribution quantiles. \\

Changes in the magnitude and the frequency of extreme events can be caused by changes in the mean values, in the variability, in the heavy-tailedness or by a combination of these factors (\cite{KatzBrown, ipcc_2012_special, lewis}). The application of non-stationary GEV distributions enables us to investigate which factors contribute to what extent at different geographic locations. In addition, we will investigate whether changes in the different distribution parameters occur simultaneously or if changes in some parameters precede changes in others. \\

Several non-stationary models based on GEV distributions have been proposed to describe the influence of climate change on climate extremes: \cite{panagoulia} proposed a GEV distribution with the parameters polynomially depending on time and showcased its application to precipitation data from Greece. In a similar way, \cite{sarhadi} constructed non-stationary models with different degrees of freedom and evaluated them using Bayesian inference and Markov chain Monte Carlo techniques. \cite{tianlisun} extended an idea first proposed in \cite{cannon} and used neural networks to choose between a variety of non-stationary models with different covariates that can interact with each other. The approach of combining GEV distributions with logistic functions gives us the possibility to investigate developments in extreme temperature over a long time span and on a global level and to research how changes in extreme temperatures will unfold in different regions. \\

The rest of this paper is organized as follows: In the next section, the temperature data set and the logistic models as well as the model-fitting algorithm are presented. The results of applying the models to the data are shown in the section thereafter. In addition to that, results of a simulation study that is conducted to investigate the accuracy of the model fitting algorithm are also discussed there. The section is followed by a discussion section, and a section on conclusions and an outlook finalize the article.

\section{Data and Methods}
\label{sec:methods}
\subsection{Data}

\begin{figure}
    \center
    \includegraphics[width=\textwidth]{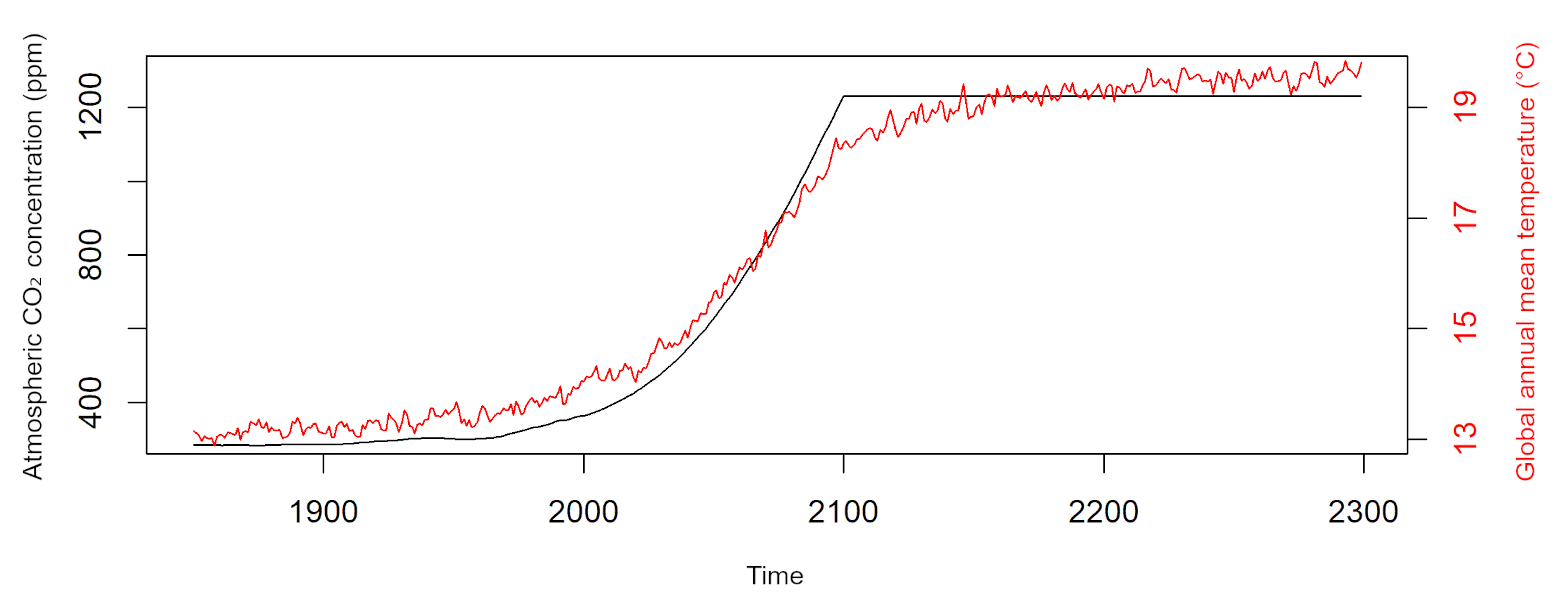}
    \caption{{\bf Atmospheric CO\textsubscript{2} concentration and global annual mean temperature.} Black line: The atmospheric CO\textsubscript{2} concentration (in ppm) that was used for the model run. The CO\textsubscript{2} concentration follows the RCP8.5 scenario (\cite{riahi}) until $2100$ and is kept constant afterwards. Red line: The annual global mean temperature (in $\DegC$) according to the AWI-ESM model run.}
    \label{fig1}
\end{figure}

We investigate monthly temperature data at two meters above surface from the AWI-ESM, a global climate model developed by the Alfred Wegener Institute in Bremerhaven, Germany. This model extends the AWI Climate Model (AWI-CM; \cite{awi-cm1,awi-cm2}) by adding dynamically coupled components for land cover (vegetation) and ice sheets. The AWI-CM is a fully coupled global climate model that consists of the atmospheric model ECHAM6 (\cite{echam}), the land surface model JSBACH2.11 (\cite{jsbach}) and the ocean-sea ice model FESOM1.4 (\cite{fesom}). The ice sheets are calculated using the dynamical ice sheet model PISM1.1 (\cite{winkelmann},\cite{martin}). The data we use consist of a simulation of the historical climate from $1850$ to $2005$ and a future simulation from $2005$ to $2300$ that follows the RCP8.5 scenario of the IPCC (\cite{riahi}). From the year $2100$ on, CO\textsubscript{2} emissions are assumed to be zero, see Fig~\ref{fig1}. \\

The data we use here were also used by \cite{ackermann} to investigate changes in ocean currents. The AWI-ESM is part of the model comparison project CMIP6 (\cite{eyring}) and has been used in several studies investigating the historical climate and future climate scenarios (\cite{shi, lohmann_butzin, niu}). \\

\subsection{The block maxima approach}

The models we create and apply in this work are based on the well-established block-maxima approach, for which we will now briefly present the theoretical foundation. For a random variable $X$ with unknown probability distribution, we investigate the distribution of the maximum of independent and identically distributed copies $X_1,\dots,X_n$ of it: $Y^{(n)} := \max_{i=1,\dots, n}(X_i)$. We assume the existence of suitable normalizing sequences $a_n > 0$ and $b_n$ such that these block-wise maxima converge in distribution as the block size $n$ tends to infinity:
\begin{eqnarray}
\label{gev_def}
\frac {Y^{(n)} - b_n} {a_n}\xrightarrow{\mathcal{D}} H. 
\end{eqnarray}
It is shown by \cite{frechet}, \cite{fisher_tippett} and \cite{gnedenko} that in this case, $H$ must follow a GEV distribution. The GEV distribution has three parameters: location ($\mu$), scale ($\sigma > 0$) and shape ($\gamma$), and its probability distribution function is given by 
\begin{eqnarray}  
F_{\mu,\sigma,\gamma}(x) = \begin{cases} \exp(-\exp(-\frac{x-\mu}{\sigma})) & \gamma = 0 \\ \exp(-\max(0,1+\gamma \frac{x-\mu}{\sigma})^{-\frac{1}{\gamma}}) &\gamma \neq 0.
\end{cases}
\end{eqnarray}
While location and scale parameters correspond very roughly to mean and standard deviation, the shape parameter is a measure of the heavy-tailedness of the distribution. Location and scale parameter of the distribution $H$ depend on the choice of $a_n$ and $b_n$, while the shape parameter does not. It is therefore justified to say that $X$ is in the domain of attraction of a unique shape parameter value $\gamma$ if Eq~\ref{gev_def} is fulfilled for some normalizing sequences $a_n$,$b_n$ and some limiting distribution $H$ with shape parameter $\gamma$. If we assume that Eq~\ref{gev_def} holds true for the underlying distribution of given data $y_1^{(n)},y_2^{(n)},\cdots$ and find suitable values $a_n$, $b_n$ and a GEV distribution approximating the data, we can include $a_n$ and $b_n$ into the estimator for $H$ and can thus estimate the distribution of $Y^{(n)}$ by a GEV distribution with unique parameters $\mu$, $\sigma$, $\gamma$. In the case of $\gamma$ being greater than $0$, the GEV distribution is also called a Fréchet distribution and is heavy-tailed (i.e. it features strong positive extremes that are markedly different from the non-extreme values). The GEV distribution for $\gamma = 0$ (Gumbel distribution) has exponential tails and the distributions for $\gamma < 0$ (Weibull distributions) have a finite right endpoint. For a more in-depth introduction to GEV distributions and the block-maxima approach, see \cite{mcneil}, Chapter 7.\\

The GEV distribution has been widely used in climatology as a model for block-wise maximized data, often applied to the yearly maxima of daily average temperature (\cite{hasan}; \cite{cueto}; \cite{kharin_zwiers_zhang}). Since our data is given in monthly intervals, we group them into blocks of three years. This way, we have a blocksize of $36$ and a total number of blocks of $150$. Note that we do not need to do an adjustment for seasonality, because in the presence of a strong seasonality, the three-yearly maxima are selected from the season with the warmest temperature anyway. \\

A requirement for applying the GEV distribution is that the underlying data be independent, which is in general not a justified assumption for climate data, since one extreme event can occur over a time span that includes a boundary between two blocks and can then be responsible for two dependent consecutive block maxima. An approach to overcome this issue is to introduce a fixed time span $\tau$ that is assumed to be a typical duration of an extreme event and then to adapt the maxima selection process in order to ensure that the selected values from the blocks are at least a time span of $\tau$ apart. Specifically, if for two consecutive blocks the block maxima are less than $\tau$ apart, the lower one of these maxima is discarded and is replaced by the block maximum based on only those values that are distant enough from the maximum of the other block. This method, which was developed first by \cite{tawn} and then employed for example by \cite{bijl} and \cite{mendez}, will also be applied to our data, using a time span $\tau$ of three months.

\subsection{Models for non-stationary GEV parameters}

\begin{figure}[!h]
    \centering
    \includegraphics[width=0.7\textwidth]{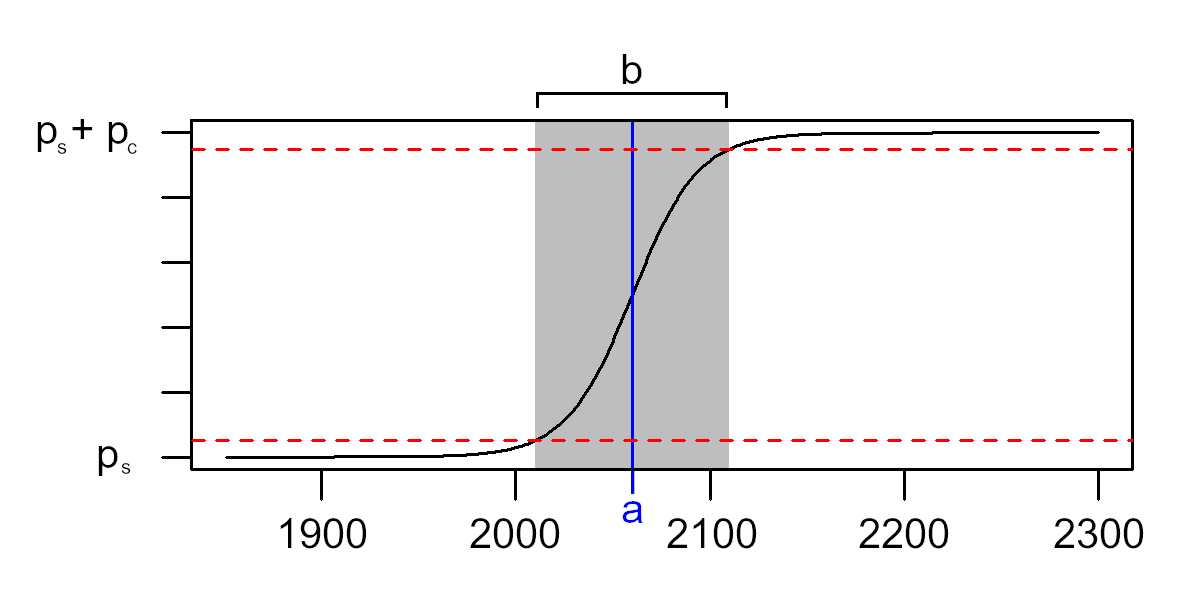}
    \caption{{\bf Visualization of the parameter values of the logistic models.} A sigmoidal curve following Eq~\ref{eq4} with parameters $a=2060$ and $b=100$ is displayed. Parameter $a$ corresponds to the time point at which half of the transition from $p_s$ to $p_s + p_c$ is completed. Ninety percent of this transition take place within the interval $[a-\frac b 2, a + \frac b 2]$, so parameter $b$ describes the approximate time span of the transition.}
    \label{fig2}
\end{figure}

To model the effects of changes in the climate, the GEV distributions we use need to have time-dependent distribution parameters. Due to the reasoning laid out in the introduction, we choose logistic functions to describe the change of the GEV parameters over time. The logistic function we use as the basis for our models is given as
\begin{eqnarray}
\label{eq:f}
    f(x) = \frac {1} {1 + \exp(-2 \cdot \log(19) \cdot x)}.
\end{eqnarray}
It describes a growth limited by $0$ for $x \rightarrow -\infty$ and by $1$ for $x \rightarrow \infty$ with the highest growth rate at $x = 0$. The constant $2 \cdot \log(19)$ in the exponential function is used for better interpretability of the parameters of the models we will present below, it ensures that $90\%$ of the change from $0$ to $1$ takes place in the interval $[-\frac 1 2, \frac 1 2]$. We use the function in our models in the following way: For each of the three GEV parameters $p \in (\mu, \sigma, \gamma)$ we describe its temporal development as
\begin{eqnarray} \hat{p}(t) = p_s + p_c \cdot f \Big{(} \frac {t-a} b \Big{)}.
\label{eq4}
\end{eqnarray}
The model parameter $p_s$ describes the "initial state" and $p_c$ describes the total magnitude of the change. The model parameters $a$ and $b$ control the timing of the change. Parameter $a$ indicates the time point at which the growth rate is highest (which is also the time point at which exactly half of the change from $p_s$ to $p_s+p_c$ is completed) and parameter $b$ indicates the approximate duration of the change (in the sense that $90\%$ of the total change takes place in the time span $[a-\frac b 2, a + \frac b 2]$). See also Fig~\ref{fig2} for a visualization. \\

This leads to the following model: \\

\textbf{Model 1a.}
The three GEV parameters location, scale and shape are described using a logistic curve, using for each parameter a different initial value and amount of change. The parameters $a$ and $b$ are the same for location, scale and shape.
\begin{align*}
\hat{\mu}(t)    &= {\mu}_{s}    + {\mu}_{c}    \cdot f \Big{(} \frac {t-a} b \Big{)} \\
\hat{\sigma}(t) &= {\sigma}_{s} + {\sigma}_{c} \cdot f \Big{(} \frac {t-a} b \Big{)} \\
\hat{\gamma}(t) &= {\gamma}_{s} + {\gamma}_{c} \cdot f \Big{(} \frac {t-a} b \Big{)}
\end{align*}

As pointed out by \cite{Schaer}, the evolution of extreme events may be different from that of mean and variance (which may show different behaviors among themselves). It may therefore be necessary to allow for changes in location, scale and shape to take place at different times and over different durations. This leads to the following more complex model: \\

\textbf{Model 1b.}
This model is the same as Model 1a, but with individual parameters $a_{\mu}$, $b_{\mu}$, $a_{\sigma}$, $b_{\sigma}$ and $a_{\gamma}$, $b_{\gamma}$ being used for location, shape and scale of the GEV distribution.
\begin{align*}
\hat{\mu}(t)    &= {\mu}_{s}    + {\mu}_{c}    \cdot f \Big{(} \frac{t-a_{\mu}} { b_{\mu}} \Big{)} \\
\hat{\sigma}(t) &= {\sigma}_{s} + {\sigma}_{c} \cdot f \Big{(} \frac{t-a_{\sigma}} { b_{\sigma}} \Big{)} \\
\hat{\gamma}(t) &= {\gamma}_{s} + {\gamma}_{c} \cdot f \Big{(} \frac{t-a_{\gamma}} { b_{\gamma}} \Big{)}
\end{align*}

When applying non-stationary GEV distributions, it is often assumed that the only time-dependent parameters are location and scale, while the shape parameters is kept constant (\cite{nogaj, panagoulia}). This approach leads us to a second type of model: \\

\textbf{Model 2a.}
The GEV parameters location and scale are described using a logistic curve, using for each parameter a different initial value and amount of change. The parameters $a$ and $b$ are the same for location and scale. The shape parameter is kept constant over the whole time interval.
\begin{align*}
\hat{\mu}(t)    &= {\mu}_{s}    + {\mu}_{c}    \cdot f \Big{(} \frac {t-a} b \Big{)} \\
\hat{\sigma}(t) &= {\sigma}_{s} + {\sigma}_{c} \cdot f \Big{(} \frac {t-a} b \Big{)} \\
\hat{\gamma}(t) &= {\gamma}_{const}
\end{align*}

\textbf{Model 2b.}
This model is the same as Model 2a, but with individual parameters $a_{\mu}$, $b_{\mu}$, $a_{\sigma}$, $b_{\sigma}$ being used for location and scale of the GEV distribution.
\begin{align*}
\hat{\mu}(t)    &= {\mu}_{s}    + {\mu}_{c}    \cdot f \Big{(} \frac{t-a_{\mu}} { b_{\mu}} \Big{)} \\
\hat{\sigma}(t) &= {\sigma}_{s} + {\sigma}_{c} \cdot f \Big{(} \frac{t-a_{\sigma}} { b_{\sigma}} \Big{)} \\
\hat{\gamma}(t) &= {\gamma}_{const}
\end{align*}

\begin{figure}[!h]
    \centering
    \includegraphics[width=0.7\textwidth]{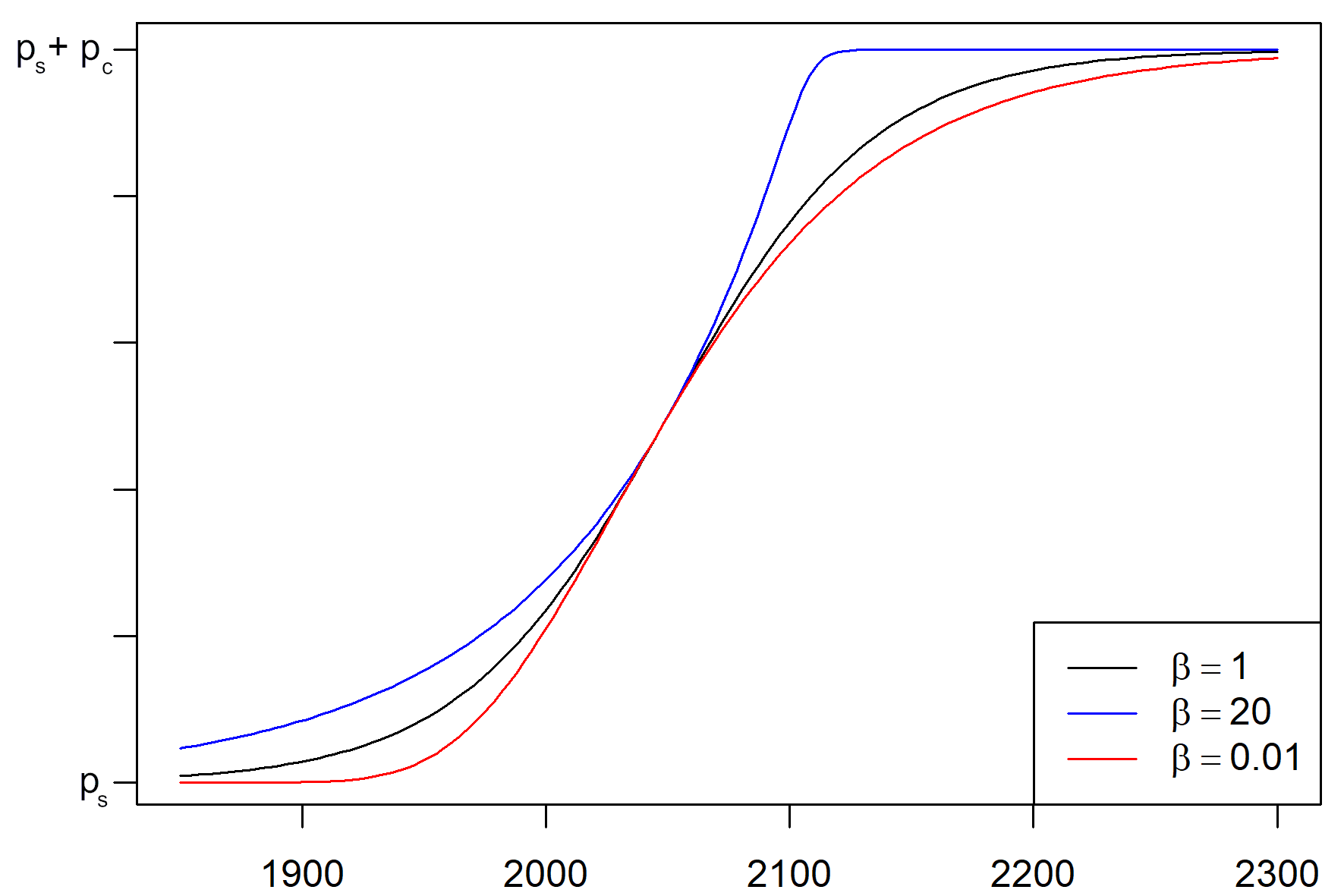}
    \caption{{\bf Visualization of the parameter $\beta$ of the Richards function.} The plot shows Richards functions $g_\beta$ for different values of $\beta$. The Richards function for $\beta = 1$ is identical to the logistic function $f(x)$. The point of the highest growth rate is shifted to the right for $\beta > 1$ and to the left for $\beta < 1$. For all lines shown, the other parameters used are $a=2050$, $b=100$.}
    \label{fig3}
\end{figure}

The logistic function, as used in the models above, has the limitation that the inflection point (the point of the strongest growth) is exactly in the middle of the curve, having always a value of $p_s + \frac 1 2 {p_c}$. To allow for more flexibility, a generalized function that was proposed by \cite{richards} can be used. For $\beta > 0$,  the Richards function is defined as

\begin{eqnarray}  g_\beta(x) = \Big{(} 1 + (2^\beta - 1) \cdot \exp\Big{(}-\log \Big{(}\frac {0.95^{-\beta}-1} {0.05^{-\beta}-1}\Big{)} \cdot x\Big{)} \Big{)} ^{ -\frac {1} {\beta}}
\end{eqnarray}

and use it to describe the time-changing GEV parameters $p \in (\mu, \sigma, \gamma)$:
\begin{eqnarray}   \hat{p}(t) = p_s + p_c \cdot g_\beta \Big{(} \frac {t-a} b \Big{)}.
\end{eqnarray}
The interpretation of the parameters $p_s$ and $p_c$ remains unchanged. The parameter $a$ describes, as before, the time point at which the model attains the midpoint of the change (the value $ p_s + \frac 1 2 p_c$). In the previous models, this was also the point of the highest growth rate, while here, the inflection point depends on the value of the parameter $\beta$. For $\beta = 1$, the model reduces to the previous model ($g_1$ is equal to $f(x)$), while the inflection occurs at a later time point than $a$ for $\beta > 1$ and at an earlier time point for $\beta < 1$. The parameter $b > 0$ controls the velocity of the change in such a way that the change from $p_s+\frac 1 {20} p_c$ to $p_s + \frac {19} {20} p_c$ ($90\%$ of the total amount of change) takes place in an interval of length $b$. Because of the asymmetry of the function $g_\beta$ for $\beta \neq 1$, this interval is no longer $[a-\frac b 2, a+\frac b 2]$, but shifted to the left for $\beta > 1$ and to the right for $\beta < 1$.  In Fig~\ref{fig3}, plots of the model function for different parameter values are shown.\\

Using the Richards function $g_\beta$ instead of $f(x)$ in the previous models gives us four models that we denote by adding the letter R to the model name. Compared to the models using the function $f(x)$, model 1aR and 2aR have one additional model parameter $\beta$, while model 1bR and 2bR feature additional model parameters for the non-constant GEV parameters $\beta_\mu, \beta_\sigma$ and (only Model 1bR) $\beta_\gamma$. \\

\subsection{Model fitting and selection}

Non-stationary GEV distributions can be fitted to data using Maximum Likelihood Estimators, see \cite{mudelsee}, Chapter 6.3 and \cite{eladlouni}. For the numerical optimization required fitting algorithm, the L-BGFS-B algorithm is used. For this purpose, the models are reparametrized to no longer use the parameters $p_c$ for $p \in \{\mu, \sigma, \gamma \}$ describing the magnitude of change, but parameters $p_e := p_s + p_c$ describing the values after the change instead. This makes it possible to ensure in an easy way that all values of $\sigma(t)$ are positive (using the condition $\sigma_e > 0$ instead of the equivalent $-\sigma_c < -\sigma_s$). To determine suitable starting values for the parameters $\mu_s$ and $\sigma_s$, a stationary GEV distribution is fitted to the first third of the data of the time series investigated, yielding estimates $\hat\mu$ and $\hat\sigma$, and starting values are selected randomly from the intervals $[\hat\mu-5, \hat\mu+5]$ and $[\max(0,\hat\sigma-5), \hat\sigma+5]$. The same is done for the parameters $\mu_e$ and $\sigma_e$ using the last third of the time series data. Since the estimation of the shape parameter is not very reliable for small samples, starting values for $\gamma_s$, $\gamma_e$ or $\gamma_{const}$ are not determined that way, but chosen randomly from the interval $[-1, 1]$. Random selection from an interval is also done for all other model parameters using suitable, large intervals to select values from. The stationary GEV distributions are fitted using the R package "EnvStats" by \cite{millard}. The optimization algorithm is run several times with different starting values in order to find a global maximum of the likelihood function. \\

To choose the best model out of the different models presented here, we apply the Bayesian Information Criterion (BIC; \cite{schwarz_bic}). In a simulation study by \cite{panagoulia}, the BIC performed better than the Akaike Information Criterion (AIC) when applied to non-stationary GEV distributions. To test the goodness-of-fit of the models, note that a 
$\mathrm{GEV}(\mu,\sigma,\gamma)$-distributed random variable can be transformed to a $\mathrm{GEV}(1,1,1)$ distribution (a so-called unit Fréchet distribution) by applying the transformation
\begin{eqnarray}
 G_{\mu,\sigma,\gamma}(z) = \max\Big(0, 1-\gamma \cdot \Big(\frac {z-\mu} {\sigma}\Big)\Big)^{- \frac 1 {\gamma}}.
 \end{eqnarray}
 
By applying $G_{\hat\mu,\hat\sigma,\hat\gamma}(z)$ with the (time-dependent) estimated model parameters to the data, we obtain for each grid point a time series that is unit Fréchet distributed if the model assumptions are true. We test the hypothesis of the transformed data being unit Fréchet distributed using a one-sample Kolmogorov-Smirnov test (\cite{stephens}). \\

\subsection{Proof of concept using simulated data}
Before applying the models to climate data, we first test how accurately model parameters can be estimated under ideal conditions. For each of the eight models presented above, we prescribe values for the model parameters, simulate data following the corresponding non-stationary GEV distribution, and fit the model to the data. We then compare the estimated model parameters with the real ones. \\

In addition to that, we test how susceptible the models that use the logistic function are to model misspecification. To this end, we simulate data from the models as above, but replacing the function $f(x)$ from Eq~\ref{eq:f} with the following three functions of a similar sigmoidal shape that are known for example as activation functions for neural networks (\cite{menon, bagul}):
\begin{eqnarray}
    g_1(x) &= \frac 1 2 + \frac 1 \pi \arctan\Big(\frac \pi 4 x\Big) \\
    g_2(x) &= \frac 1 2 + \frac x {4 \sqrt{1+\frac {x^2} 4}} \\
    g_3(x) &= \frac 1 2 + \frac 1 2 \mathrm{erf}\Big(\frac {x\sqrt{\pi}}  4\Big).
\end{eqnarray}
The simulated data follow a non-stationary GEV distribution with time-dependent distribution parameters $\mu_t$, $\sigma_t$, $\gamma_t$. We then calculate estimates $\hat\mu_t$, $\hat\sigma_t$, $\hat\gamma_t$ by fitting the original model (using function $f(x)$) to the data, and calculate the squared difference of given and estimated GEV parameters, integrated over time 
\begin{eqnarray}
    \int_{t \in T} (p_t-\hat{p}_t)^2 \mathrm{d} t 
\end{eqnarray}
for the three GEV parameters $p \in \{\mu, \sigma, \gamma\}$. 

\section{Results}
\label{sec:results}
\subsection{Results of the simulation study}
We simulated $1000$ time series for each model using the parameters $\mu_s=20$, $\mu_c=10$, $\sigma_s=2, \sigma_c=1$. The parameters for the shape parameter were $\gamma_s=0.1$ and $\gamma_c=0.1$ (models with a varying shape parameter) or $\gamma_{const} = 0.1$ (models with a constant shape parameter). For the models describing a simultaneous change in all parameters we used the parameters $a=2075$, $b=30$, otherwise we used $a_\mu=2050$, $b_\mu=30$, $a_\sigma=2075$, $b_\sigma=30$ and, if applicable, $a_\gamma=2100$ and $b_\gamma=30$. The models based on the Richards function instead of the logistic function additionally had a parameter $\beta$ (or parameters $\beta_\mu$, $\beta_\sigma$, $\beta_\gamma$, respectively) equal to $5$. \\

\begin{figure}[!h]
    \centering
    \includegraphics[width=0.6\textwidth]{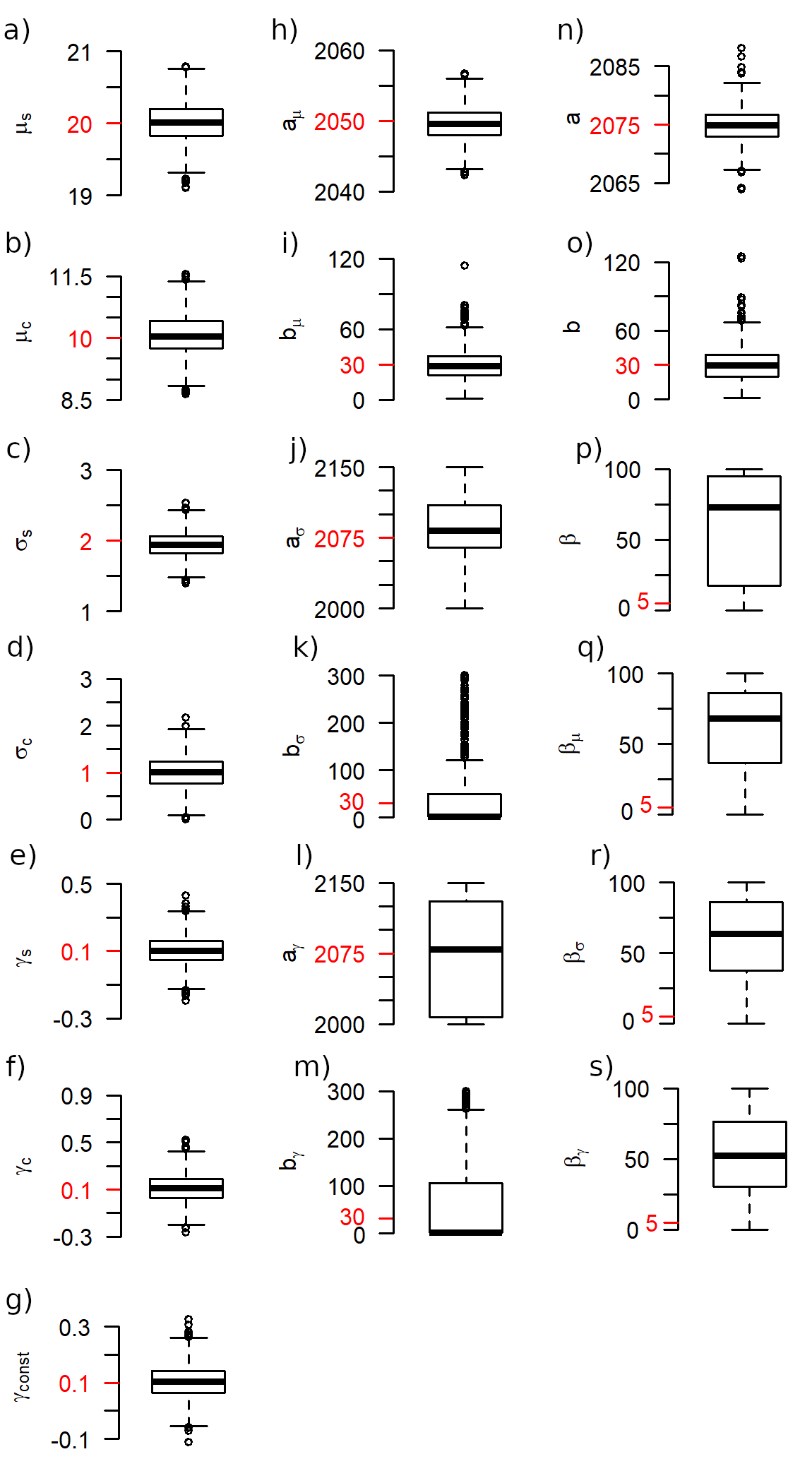}
    \caption{{\bf The accuracy of the parameter estimation for simulated data.} The accuracy of the maximum likelihood estimators is investigated by applying the models to data that were generated following the respective model. For each parameter, a boxplot of the estimates is shown, with the real parameter value indicated in red. Since the results for each parameter are very similar across the models, only one boxplot is presented per parameter. The model depicted is 1a (a-f, n, o), 1b (h-m), 2a (g), 1aR (p) and 1bR (q-s). }
    \label{fig4}
\end{figure}

It turned out that for each parameter, the estimation quality is similar for all models in which the parameter occurs. In particular, the estimation is not more inaccurate for the more complex models with a higher number of parameters. For each parameter, boxplots of the estimates are shown in Fig~\ref{fig4}. Since the estimates are similar for each model, only the boxplot for one model per parameter is shown. As can be observed, the start and change values of the GEV parameters are in general well estimated, and the same is true for the parameter $a$ and $a_\mu$ if they exist in the model. The estimates for parameters $b$ and $b_\mu$ are in most cases close to the real value, but there are also some cases of a considerable misestimation (with a real parameter value of $30$, the estimates take values of up to $120$). The parameters describing a separate change in scale, $a_\sigma$ and $b_\sigma$, are estimated much worse than the other ones, estimates that are far away from the original value occur regularly. In addition, parameter $b_\sigma$ is in most cases underestimated, with the median of the estimates being far lower than the real value, while cases of a strong overestimation of this parameters also occur. The same can be said for the parameters $a_\gamma$ and $b_\gamma$, but their estimation accuracy is even lower. \\

The estimation of the additional $\beta$ parameters that appear in the models using the Richards function turned out to be very problematic for all models. The estimated values are usually far away from the real ones, and even the medians of the estimates are between $50$ and $75$ and not even close to the real parameter values of $5$. A reliable estimation of the $\beta$ parameter of the Richards function seems to be impossible in general using the method we employed here. Because of that, the models using the Richards function will not be considered further and only the models using the logistic function will be applied to the data. It was considered also to exclude model 1b because of the high estimation inaccuracy of the parameters $a_\gamma$ and $b_\gamma$, but for the sake of completeness, the model was kept. In the next section it will be seen that this model is rarely favored by the BIC anyway. \\

The results of the simulation study investigating model misspecification due to other logistic functions than $f(x)$ are shown in Table~\ref{table1}. Results are shown only for model 1a, but are similar for the other logistic models. It can be observed that for data that were created using one of the functions $g_i(x)$, the errors are similar to those using function $f(x)$. Therefore, model misspecification caused by the usage of different sigmoidal functions does not have a strong negative impact on the estimation accuracy and there is no need for using different functions than $f(x)$ when applying the models to the data. 

\begin{table}[!ht]
    \centering
    \begin{tabular}{|c+c|c|c|}
\hline
    {\bf Function used} & {\bf Location} & {\bf Scale} & {\bf Shape} \\
    \hline
    $f(t)$   & 0.232 & 0.057 & 0.008 \\ \hline
    $g_1(t)$ & 0.230 & 0.060 & 0.008 \\ \hline
    $g_2(t)$ & 0.278 & 0.060 & 0.008 \\ \hline
    $g_3(t)$ & 0.235 & 0.060 & 0.008 \\ \hline
    \end{tabular}

\caption{{\bf The influence of model misestimation on the estimation accuracy.} The squared difference of constructed and estimated GEV parameters is shown for the GEV parameters location, scale and shape. The data were simulated using the sigmoidal function in the left-most column of the table while the model that was fitted to the data always uses the function $f(t)$. The model used is model 1a, results for the other models are similar. The errors are averaged over $5000$ iterations. }
    \label{table1}
\end{table}

\subsection{Application to the data}

\begin{figure}[!h]
    \includegraphics[width=\textwidth]{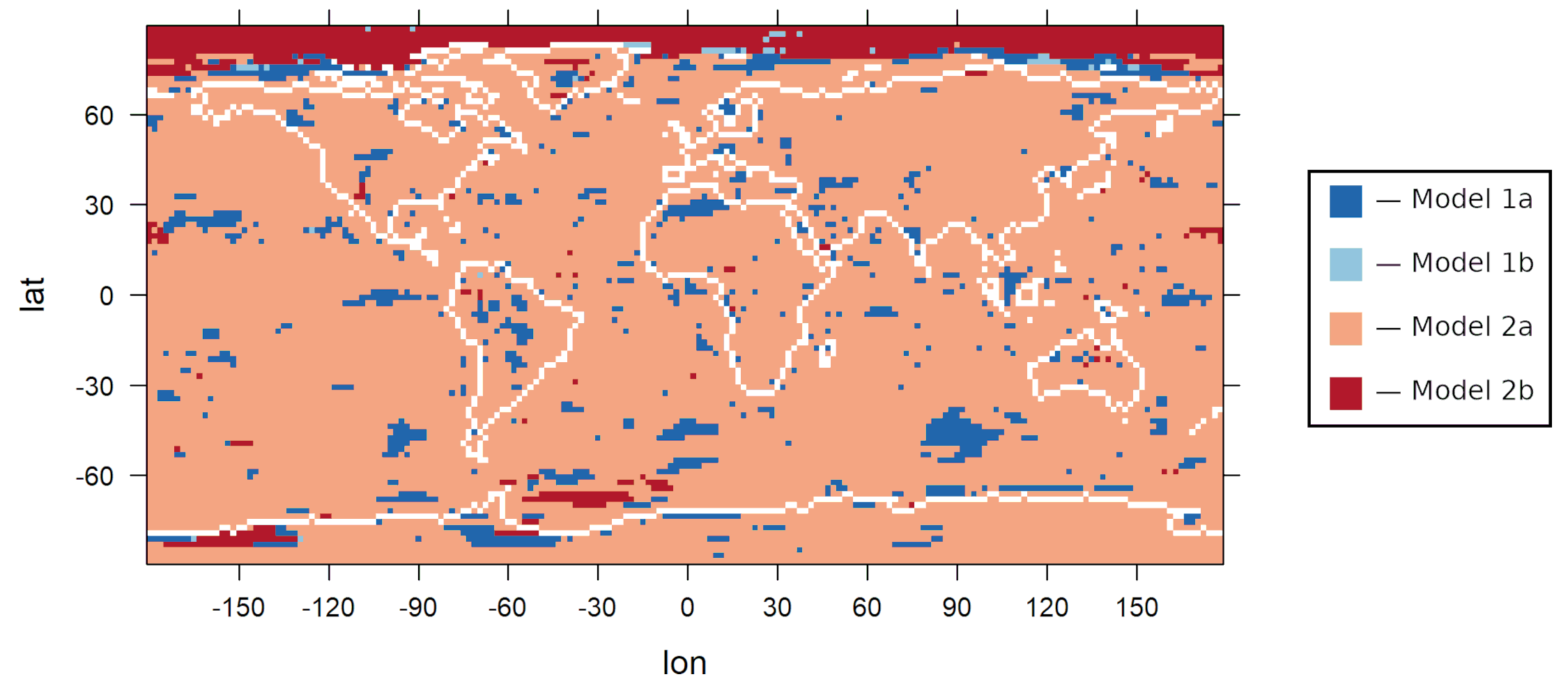}
    \caption{{\bf The preferred model according to the Bayesian Information Criterion for each grid point.} The logistic models 1a, 1b, 2a and 2b are applied to three-yearly maxima of monthly temperature data and the BIC is used to determine the optimal one out of these for each grid point.}
    \label{fig5}
\end{figure}

As mentioned before, the only models that are applied to the data are the four models based on the logistic function. In Fig~\ref{fig5}, the best model according to the BIC is shown for each grid point. It can be observed that the models with a constant shape parameter (model 2a and 2b) are often preferred over the models with a varying shape parameter; one of these models is selected for $92.4\%$ of all grid points. There are many smaller regions in which a model with a varying shape parameter is preferred instead, a clear connection between those regions could not be identified. On the other hand, a pattern is visible regarding the question whether a model with simultaneous changes in location and scale (and, if applicable, shape) parameter is selected or not: Models with individual change parameters for the different GEV parameters are preferred almost exclusively in high-latitude regions, and in particular throughout the whole region around the North Pole from ca. $80^\circ \textrm{N}$ onward. In all other regions models with a simultaneous change in the GEV parameters are favored.\\

\begin{figure}[!h]
    \includegraphics[width=\textwidth]{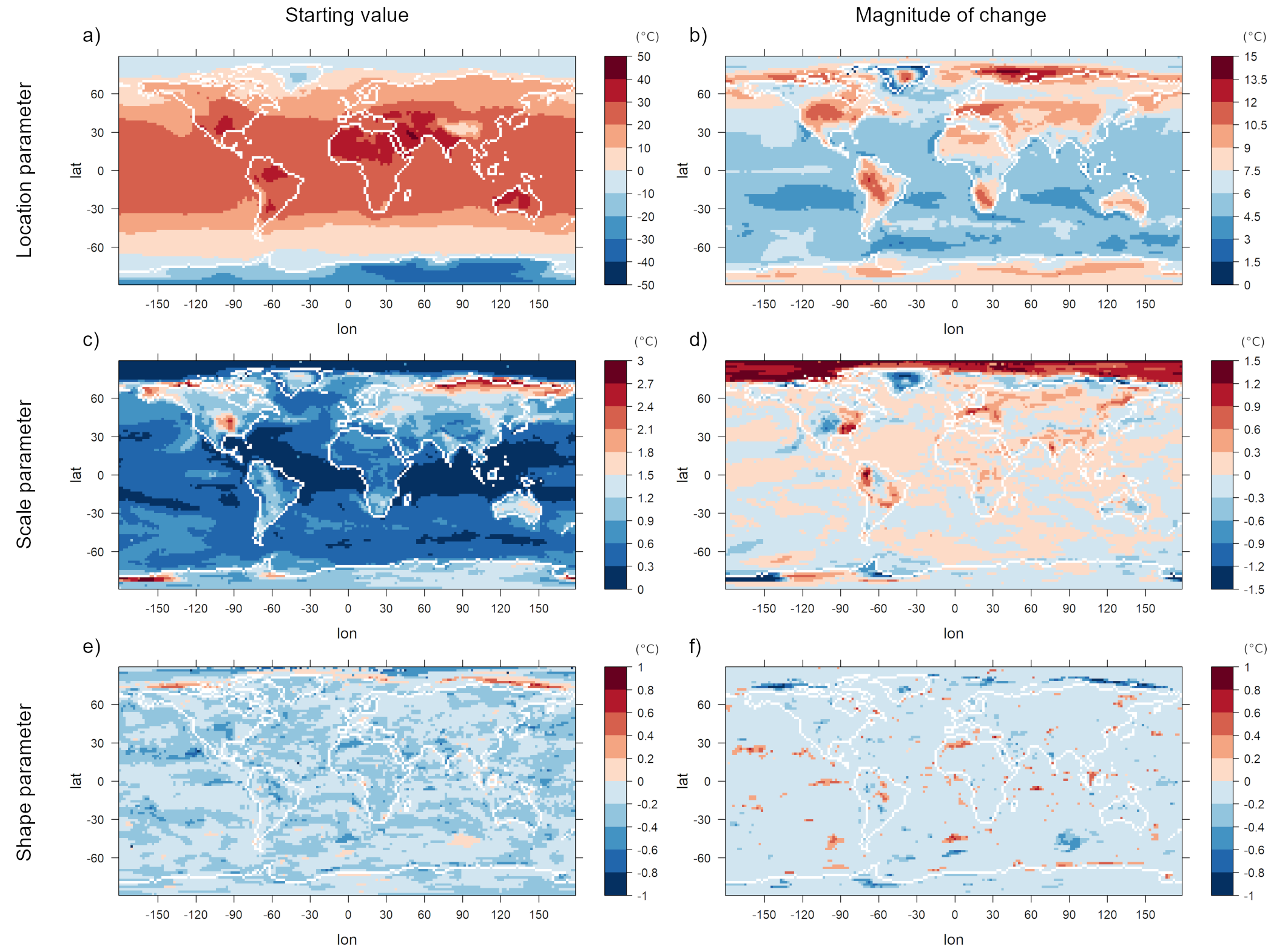}
    \caption{{\bf The estimates for the starting values and the amounts of change of the GEV parameters.} The estimates for the parameters $\mu_s$ (a), $\sigma_s$ (c), $\gamma_s$ (e) and $\mu_c$ (b), $\sigma_c$ (d), $\gamma_c$ (f), describing starting value and total amount of change over time of the GEV parameters location, scale and shape. The models are applied to three-yearly maxima of monthly temperature data. For each grid point, the estimates of the model that was preferred by the Bayesian Information Criterion are shown. Units are $\DegC$.}
    \label{fig6}
\end{figure}

All models share the parameters $\mu_s, \mu_c, \sigma_s$ and $\sigma_c$ describing the starting value and the magnitude of change of the location and scale parameters. Parameters $\gamma_s$ and $\gamma_c$ are estimated only for Models 1a and 2a, but for the other models we can define $\gamma_s$ as the constant estimate of the shape parameter and $\gamma_c$ as equal to zero. Using this definition, the values of the six parameters are shown in Fig~\ref{fig6}. For each grid point, we show the estimates of the model that is preferred by the BIC at that grid point. \\

As expected, the starting value of the location parameter depends highly on the latitude and the climate zone of the grid point investigated. The starting values of the scale parameter show a dependency on the continentality of the climate: the scale parameter is lowest over the oceans and highest in the very continental region of Siberia. High values of the scale parameter can also be observed in some smaller regions in North America, Alaska and Antarctica. The starting values of the shape parameter are quiet homogeneous, attaining mostly slightly negative values that indicate that no strong positive extremes are present. The only exception to this are some regions in the Arctic Ocean, north of the regions with the high scale parameter discussed above. In these regions, the high values of the shape parameter together with low values of the scale parameter indicate a climate that is in general fairly homogeneous, but with occasionally strong outliers. \\

Investigating now the parameters describing the magnitude of change in the GEV parameters, we observe strong changes in the location parameter especially over land masses, with an increase of $10 \DegC$ or more occurring in Europe, South Africa and the Americas in some regions. Also over all other continental land masses, changes of at least $5 \DegC$ can be observed. Over the ocean, the changes are in general much smaller. The region with the highest change in the location parameter, however, is the Arctic Ocean, showing an increase in the location parameter of up to $15\DegC$. \\

The scale parameter remains mostly the same over the oceans, while over land in most regions a slight increase can be observed. This increase is strongest in Europe, the eastern part of the United States and some regions in South America. In a few regions, most notable the western United States and Greenland, a decrease in scale occurs. Also the region in Antarctica that has a high starting value of the scale parameter sees a strong decrease in scale over time. The most striking change, however, takes place in the region around the North Pole in form of a very strong increase in the scale parameter. \\

For the shape parameter, a strong decrease is observable for the regions in the Arctic Ocean that have a high starting value for this parameter, resulting in final values of the shape parameter similar to those of the surrounding areas. Besides that, many smaller regions with high increases or decreases of the shape parameter can be seen, with no clear geographical pattern recognizable. As already seen by the results of the BIC, in most regions models with no change of the shape parameter are preferred. \\

\begin{figure}[!h]
    \includegraphics[width=\textwidth]{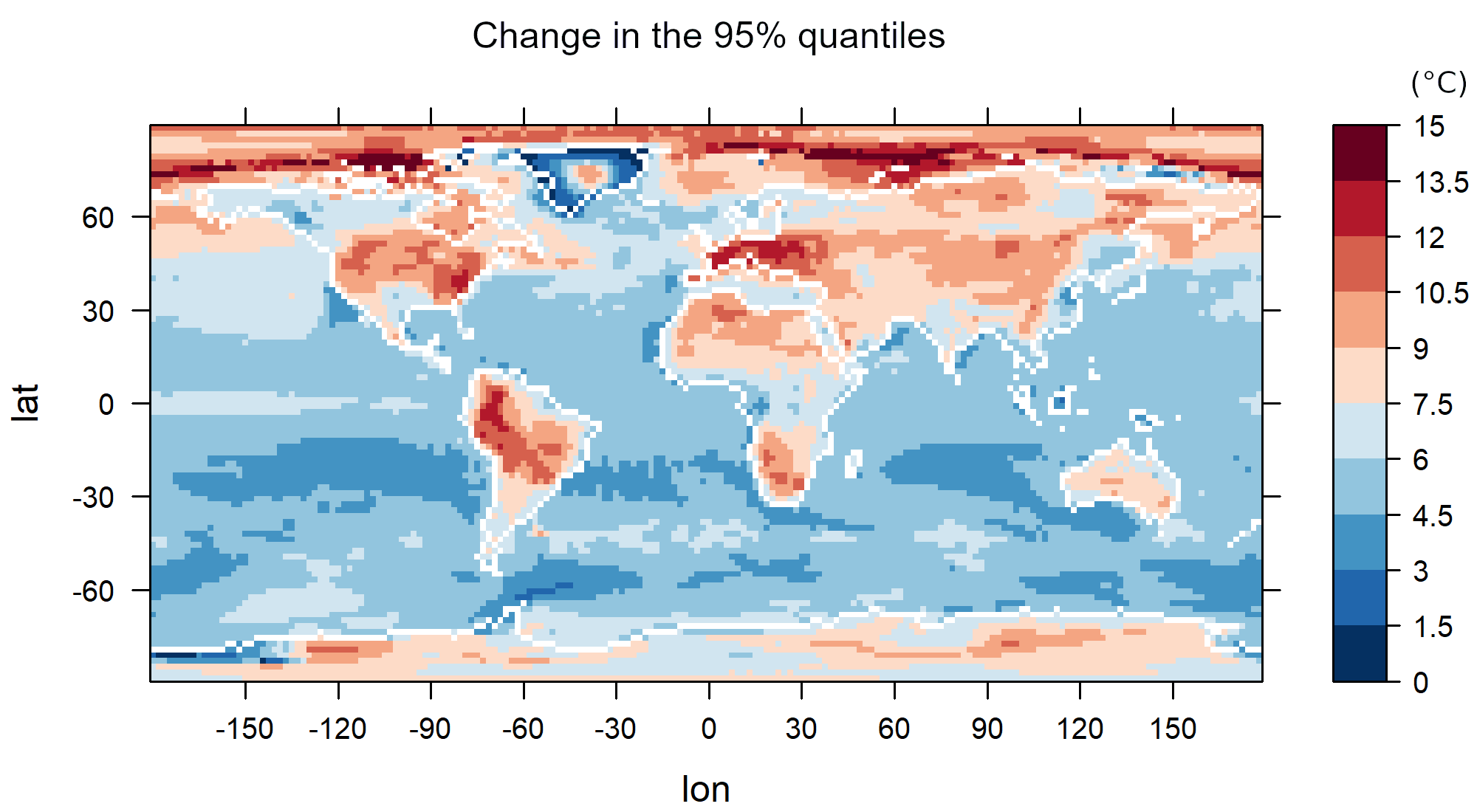}
    \caption{{\bf Changes in the $95\%$-quantiles.} For each grid point, the change in the $95\%$-quantile of the GEV distribution over time is shown. The GEV distributions are estimated by fitting the logistic models to three-yearly maxima of monthly temperature data. For each grid point, the model that is preferred by the Bayesian Information Criterion is used. Units are $\DegC$.}
    \label{fig7}
\end{figure}

To illustrate how these changes in the different GEV parameters translate into changes in extremes, in Fig~\ref{fig7}, the changes of the $95\%$ quantiles of the three-year maxima of monthly temperature are depicted. Since a shift in the location parameter of a GEV distribution directly implies an equal shift in the quantiles, it is not surprising that the figure looks similar to panel B of Fig~\ref{fig6} showing the change in location over time. However, a strong increase in the quantiles can also be seen in the region around the North Pole, in part due to changes in location and in part due to changes in scale. Taking this together, we observe a strong increase in the $95\%$ quantiles over the whole arctic region (with the exception of Greenland).\\

\begin{figure}[!h]
    \includegraphics[width=\textwidth]{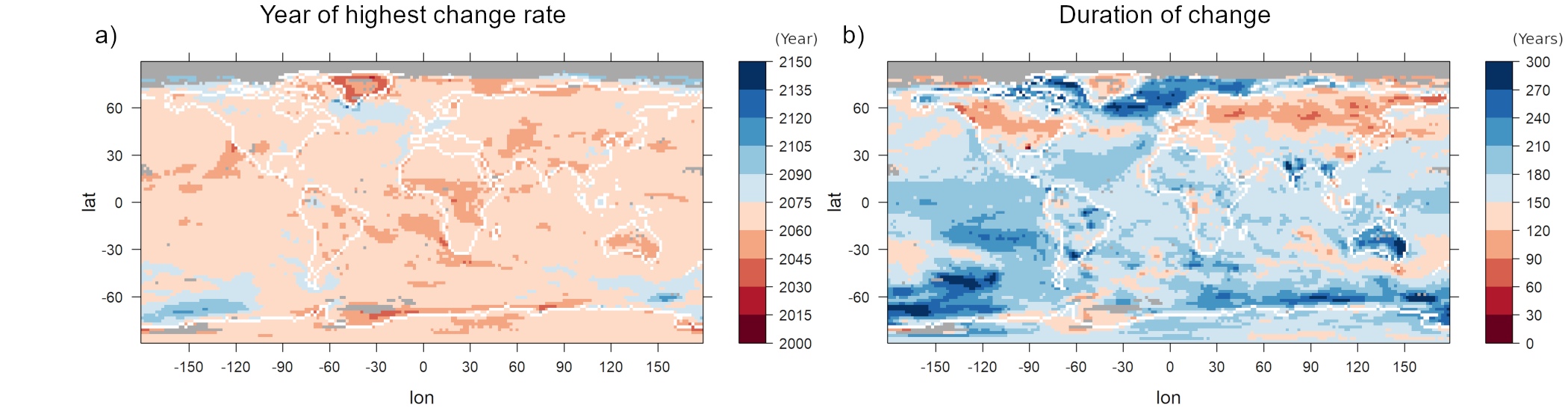}
    \caption{{\bf The estimates for timing (a) and duration (b) of change, shown for models with a simultaneous change in all GEV parameters.} The models are applied to three-yearly maxima of monthly temperature data. For each grid point, the estimates of the model that was preferred by the Bayesian Information Criterion are shown. If the preferred model at a certain grid point does not feature parameters for simultaneous changes in the GEV parameters, the grid point is grayed out. Units are years.}
    \label{fig8}
\end{figure}

We now turn our attention to the parameters describing at which time the changes take place. Models 1a and 2a have one parameter describing the time of change and one parameter describing its duration that are used for all three GEV parameters simultaneously. In Fig~\ref{fig8}, these parameters are depicted. As before, for each grid point, the estimates of the model that was favored by the BIC are shown. If the selected model at a grid point is not one of model 1a or 2a, the grid point is grayed out. It can be observed that the time around which the change takes place is consistently between $2045$ and $2075$ for most regions of the world. Exceptions are Greenland, which shows earlier changes, and some regions over the ocean in which changes occur later. The durations of the change are more varied, with a general tendency for slower changes over the ocean and rather rapid changes over land, especially in the mid-latitudes of the Northern Hemisphere. \\

\begin{figure}[!h]
    \includegraphics[width=\textwidth]{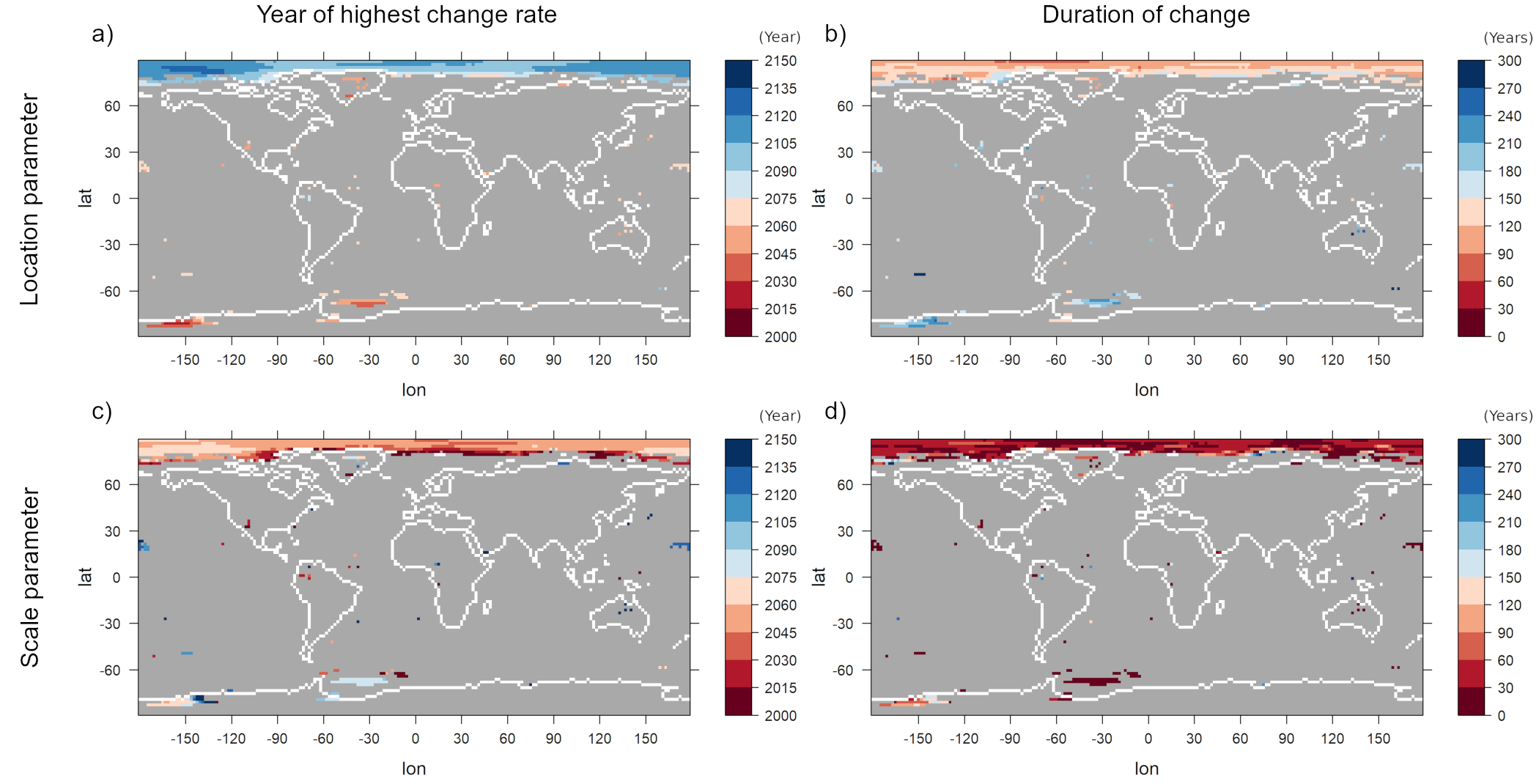}
    \caption{{\bf The estimates for timing (a, c) and duration (b, d) of change for location (a, b) and scale (c, d) parameter, shown for models with separate changes in the GEV parameters.} The models are applied to three-yearly maxima of monthly temperature data. For each grid point, the estimates of the model that was preferred by the Bayesian Information Criterion are shown. If the preferred model at a certain grid point does not feature parameters for separated changes in the different GEV parameters, the grid point is grayed out. Panels showing changes in the shape parameter are omitted since the only model featuring these parameters is selected only in very rare cases by the BIC. Units are years.}
    \label{fig9}
\end{figure}

The other two models feature individual change parameters for the location and the scale (and, in the case of Model 1b, for the shape) parameter. These values are depicted in Figure \ref{fig9} for the grid points at which one of those models is chosen. The change parameters for the shape are omitted since Model 1b is preferred over the others only for very few, isolated grid points. At all grid points, strong differences between the parameters corresponding to location and those corresponding to scale can be seen, explaining why models that allow for individual changes in the different parameters perform better there. We focus on the largest region that is not grayed out, the area around the North Pole. In this region, changes in the scale parameter take place much earlier than those in the location parameter ($2000$-$2075$ compared to $2090$-$2150$), and the scale parameter also changes considerably more rapidly than the location parameter (a duration of change of $0$-$60$ years compared to $90$-$150$ years). \\

\begin{figure}[!h]
    \centering
    \includegraphics[width=0.8\textwidth]{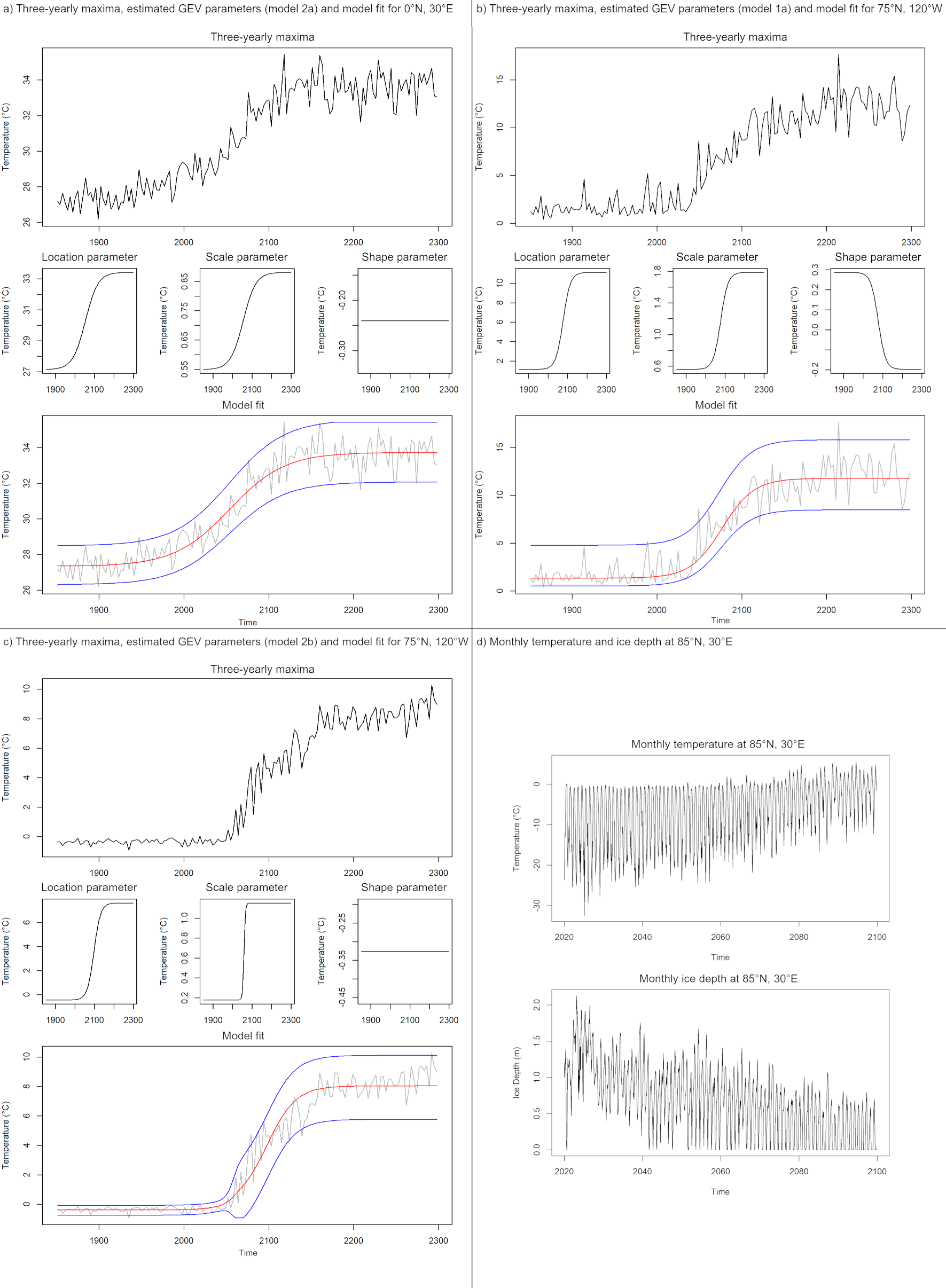}
    \caption{{\bf Detailed examination of data and fitted models at selected grid points.} (a) The three-yearly maxima of monthly temperature data at grid point $0^\circ$ N,  $30^\circ $E, together with the non-stationary GEV parameter estimates of the preferred model at this grid point and the median of the estimated distribution (red line) as well as the $95\%$ confidence interval (blue lines). (b) The same analyses as in panel a for grid point $75^\circ$ N, $120^\circ$ W. (c) The same figures as in panel a for grid point $85^\circ$ N, $30^\circ$ E. (d) The monthly temperature data and monthly ice depth for grid point $85^\circ$ N, $30^\circ$ E.}
    \label{fig10}
\end{figure}

To illustrate the models further, in Fig~\ref{fig10}a, \ref{fig10}b and \ref{fig10}c, for three grid points the three-yearly maxima of the temperature data are shown together with the estimated GEV parameters of the preferred model at that grid point and the median and $95\%$ quantiles of the estimated time-dependent GEV distributions. The grid points are chosen such that each panel features one of the three more common models: One grid point is located at the equator, the model selected there is Model 2a. The second grid point is north of Alaska in the Arctic Ocean, showcasing the application of Model 1a. The third point is located near the North Pole to give an example of a point where Model 2b is preferred. \\

A first visual inspection indicates that the models seem to fit the data reasonably well. In Fig~\ref{fig10}a, a clear logistic shape is visible in the data. In addition to the change in the location parameter indicating a rise in overall temperatures, we can detect a small increase in the scale of the distribution, corresponding to a growing variability. In Fig~\ref{fig10}b, the shape parameter shows a strong decrease, with an increase in scale at the same time. In the time series presented in Fig~\ref{fig10}c, changes in scale precede the changes in location: For a long time, the variability of the data is very low, and then increases rather suddenly around the year $2500$. Changes in the location parameter take place more slowly and the highest change rate is reached only in $2100$. In this last example, the data seem to be bounded from above by $0\DegC$ before the year $2050$. A closer inspection of the original, monthly data confirms that while the winter temperatures gradually increase, the summer temperatures do not exceed $0\DegC$ for a long time. This is due to the continued presence of ice at this location: while ice melts, its temperature is kept at $0\DegC$, and it can be seen by comparing monthly temperatures with ice depths (Fig~\ref{fig10}d) that once the ice completely melts in summer, the summer temperatures do exceed $0\DegC$ regularly. From that point on, the variability of the three-yearly maxima is strongly increased and comparable to that of other regions over land. Since changes in the location continue for a long time after the ice has melted, is becomes clear why different change parameters have to be used for location and scale for this grid point.\\

The goodness of fit of the models is tested using a Kolmogorov-Smirnov test at significance level $5\%$, which is applied for each grid point to the results of the model that is preferred at that grid point by the BIC. There is no grid point for which the hypothesis of the data following the modeled non-stationary GEV distribution is rejected. The lowest p-value obtained is $6.2\%$. It is important to keep in mind that the non-rejection of the hypothesis doesn't mean its confirmation, but still, this result is a promising indicator for the applicability of the models.

\section{Discussion}
\label{sec:discussion}

We presented mathematical models for extreme temperature that were applied to global climate data that span several hundred years and are influenced by climate change. While it is not a new approach to use non-stationary GEV distributions to investigate the development of climate extremes, most studies assume a dependency of the GEV distribution parameters on time that is either linear/polynomial (\cite{casati, kharin_zwiers, sarhadi, mahajan}) or exponential (\cite{hanel}). Consequently, the models are usually applied to short-term data. If the goal is to investigate changes in extremes on a longer time scale, the time frame is usually split up into several intervals of short length and stationary GEV distributions are fitted to each one. Then, their parameters are compared. This approach was used for global precipiation by \cite{russo}, for precipitation and temperature in Australia by \cite{perkins} and for summer temperature in the United States by \cite{hogan}. In a study by \cite{slater}, annual maxima of daily temperature data from several CMIP6 models were investigated and stationary GEV distributions were fitted to the data at different time intervals. The results are mostly in line with the results of this study, even though we used monthly and not daily data and also used a different greenhouse gas emission scenario (RCP8.5 vs. SSP370). In both studies, it was noted that the location parameter changes strongly over land and that this contributes to a large extent to the changes in extremes. A large increase in the shape parameter over the Arctic was observed by \cite{slater} together with a decrease of the scale parameter. In contrast to that, a decrease in scale parameter was visible in our study only in some parts of the Arctic. For the scale parameter, \cite{slater} identified a tendency for an increase over time in the tropics and a decrease over time in high-latitudes, a pattern that could not be found using our models. Using the approach of fitting stationary distributions to different short time intervals, it is more difficult to make statements regarding the temporal aspects of the changes. In this regard non-stationary GEV distribution are advantageous, although it is often difficult to find suitable parametrizations for the non-stationary parameters. Our approach of combining logistic functions with GEV distributions to describe climate extremes has not been used before to our knowledge. Logistic functions have, however, been used to describe historical CO\textsubscript{2} in many countries (\cite{mingmeng, koene}) and have also been applied to future projections of greenhouse gas concentrations (\cite{perez_suarez}). Climate change is closely connected to CO\textsubscript{2} concentrations, and the mean global temperature has been shown to be in an approximately linear relation to them (\cite{matthews}). This supports the idea of using logistic functions to describe extreme events under climate change as well.\\

It needs to be emphasized, though, that logistic functions are suitable for the modeling of future climate only under the condition of a cessation of greenhouse gas emissions in the future. For model data that are based on other scenarios, different functions have to be used, although logistic functions might also be useful for describing data that show a continuously rising trend in the extremes. In particular, the extraction and storage of atmospheric CO\textsubscript{2} in order to revert some consequences of climatic changes (and to prevent others) are more and more discussed. This is reflected by the SSP scenarios (replacing the RCP scenarios) used in the newer IPCC reports (\cite{IPCC_2021}), of which some predict a reduction of the atmospheric CO\textsubscript{2} levels starting in the second half of the century. A possible extension of the logistic models for such a scenario is based on the double logistic function
\begin{eqnarray}    p_{s}    + p_{c,1}    \cdot f \Big{(} 2 \cdot \log(19) \cdot \frac{t-a_{p,1}} { b_{p,1}} \Big{)} + p_{c,2}    \cdot f \Big{(} 2 \cdot \log(19) \cdot \frac{t-a_{p,2}} { b_{p,2}} \Big{)}.
\end{eqnarray}
In this formula, two logistic function are combined, allowing for the description of a change from one state to another that is not completed, but instead reverted mid-way and that finally settles on an intermediate value. \\

\section{Conclusions}
\label{sec:conclusions}
In this work, we created and applied mathematical models for the long-term development of temperature extremes that allowed us to investigate the magnitude and the timing of the changes in temperature extremes. In addition, the models differentiated between changes in the mean, the variability and the distributional shape. We summarize the conclusions of our work in the following main points: \\

1. Different logistic models were presented, and it was shown using a goodness-of-fit test that they are indeed applicable to the temperature data investigated. It was also shown in the simulation study that the model fitting algorithm is capable of estimating the model parameters with a satisfactory precision (and other models for which this was not the case were rejected). The quality of the parameter estimation does not decrease if the number of parameters increases. In addition to that, model results do not depend strongly on the choice of the underlying S-shaped function. \\

2. Depending on the geographic location, different models were favored by the BIC and different parameter values were estimated, showing a complex pattern of global warming on the climate system. The models with few parameters were favored much more often than those with a high number of parameters and a higher model complexity. In general, when analyzing future developments in extreme temperature, it needs to be kept in mind that these developments vary strongly depending on the geographic location in terms of magnitude and timing of the changes. \\

3. A strong increase in the $95\%$ quantiles of the three-year maxima could be observed in most regions of the world. In these regions, extremes will continue to rise and reach unprecedented strengths in the future. This is true especially over continents and corresponds to the well-known fact that global warming is stronger over land than over the oceans or in coastal regions (\cite{byrne, charlotte}). A particularly strong warming can also be observed over the Arctic region. This phenomenon is known for mean values under the term Arctic Amplification (\cite{serreze, cai}), here we can observe that it applies to extremes as well. \\

4. The changes in the quantiles are to a large extent due to changes in the location parameter of the GEV distribution, but in many regions also an increase in scale (variability) contributes to the higher frequency of extremes. This is in line with results of investigations of climate model data using probability ratios (\cite{vanderwiel}). The shape of the GEV distribution, an indicator of its heavy-tailedness, is seen not to change in most areas. However, in the Arctic region, a strong change in variability can be observed that is due to the melting of sea ice and that has a strong contribution to the increase of extremes in that region. Additionally, there are regions with a decrease in heavy-tailedness and an increase in scale, indicating a change from a climate that is in general homogeneous but with some strong outliers to a climate with less outliers but a higher general variability. \\

5. The changes in location and scale parameters are predicted to take place mostly simultaneously and to reach the highest rate of change in the time between $2045$ and $2075$. On the other hand, the velocity of the change differs considerably among different regions, and a particularly quick change is predicted for North America, Europe and the northern parts of Asia. In these regions, large changes in temperature are projected to take place over the course of about $100$ years. The only larger region where non-simultaneous changes in the parameters are predicted is the region around the North Pole, where changes in the location continue to take place for a long time after an abrupt increase in scale around the year $2050$.\\

\section{Outlook}
In this work, the focus was put on the univariate analysis of temperature extremes: for each grid point, the time series of temperature data was investigated separately from all others. For a better understanding of climate extremes it is important to also investigate multivariate distributions. For example, climate extremes that take place simultaneously over a large region are especially problematic because of high damages for economies and possible difficulties in providing necessary medical or humanitarian aid. Possible spatio-temporal models for climate extremes are max-stable models that are investigated for example in \cite{ribatet, dehaan} and \cite{dombry}. Since max-stable models require the data to be spatially stationary, i.e. that the joint distribution of two sites depend only on their geographical distance, they are usually only applied to small regions where such an assumption may be justified. Another approach to investigating spatial relations between climate extremes is the application of a clustering algorithm that identifies regions with similar extremal behavior \cite{contzen}. We plan to use the results in this work as a basis for the application of multivariate models in a future paper.

\section{Acknowledgements}
The authors wish to thank Manfred Mudelsee for constructive discussions and helpful suggestions. We are also grateful to Lars Ackermann for providing the temperature data and for support with uploading it to the repository.

\end{document}